\documentclass{iaus}
\usepackage{graphicx}

\title[Star distribution in open clusters]
      {The spatial distribution of stars \\
       in open clusters}
\author[N\'estor S\'anchez \& Emilio J. Alfaro]
       {N\'estor S\'anchez  \and Emilio J. Alfaro}
\affiliation{Instituto de Astrof\'{\i}sica de Adaluc\'{\i}a, CSIC, \\
             Apdo. 3004, E-18080, Granada, Spain \\
             email: {\tt nestor,emilio@iaa.es}}
\pubyear{2010}
\volume{266}
\jname{Star Clusters: Basic Galactic Building \\ 
       Blocks Throughout Time And Space}
\editors{R. de Grijs \& J. Lepine, eds.}

\begin{document}

\maketitle

\begin{abstract}
In this work we study the internal spatial structure
of 16 open clusters in the Milky Way spanning a wide
range of ages. For this, we use the minimum spanning
tree method (the $Q$ parameter, which enables one to
classify the star distribution as either radially or
fractally clustered), King profile fitting, and the
correlation dimension ($D_c$) for those clusters with
fractal patterns. On average, clusters with fractal-like
structure are younger than those exhibiting radial star
density profiles. There is a significant correlation
between $Q$ and the cluster age measured in crossing
time units. For fractal clusters there is a significant
correlation between the fractal dimension and age. These
results support the idea that stars in new-born clusters
likely follow the fractal patterns of their parent
molecular clouds, and eventually evolve toward more
centrally concentrated structures. However, there can
exist stellar clusters as old as $\sim 100$ Myr that
have not totally destroyed their fractal structure.
Finally, we have found the intriguing result that
the lowest fractal dimensions obtained for the open
clusters seem to be considerably smaller than the
average value measured in galactic molecular cloud
complexes.
\end{abstract}

\firstsection

\section{Introduction}

The hierarchical structure observed in some open clusters is
presumably a consequence of its formation in a turbulent medium
with an underlying fractal structure (\cite[Elmegreen \& Scalo
2004]{Elm04}). Otherwise, open clusters having central star 
concentrations with radial star density profiles likely reflect
the dominant role of gravity, either on the primordial gas
structure or as a result of a rapid evolution from a more 
structured state (\cite[Lada \& Lada 2003]{Lad03}). Therefore,
the analysis of the distribution of stars may yield information
on the formation process and early evolution of open clusters.
It is necessary, however, that this kind of analysis is done by
measuring the cluster structure in an objective, quantitative,
as well as systematic way. Here we study the internal spatial
structure in a sample of 16 open clusters spanning a wide range
of ages.

\section{Procedure}

\begin{enumerate}
\item We first used VizieR (\cite[Ochsenbein et al. 2000]{Och00})
to search for catalogs containing both positions and proper motions
of stars in open cluster regions.
\item We applied a robust non-parametric method to assign cluster
memberships (\cite[Cabrera-Ca\~no \& Alfaro 1990]{Cab90}). This
method makes no a priori assumptions about cluster and field star
distributions.
\item We fitted \cite[King (1962)]{Kin62} profiles to the radial
density distribution of cluster members. From these fits we
obtained both the core radius ($R_c$) and the tidal radius ($R_t$).
\item Then, we used the minimum spanning tree technique (see
Fig.~\ref{mst}) to calculate the dimensionless parameter $Q$
(see details in \cite[Cartwright \& Whitworth 2004]{Car04} and
\cite[Schmeja \& Klessen 2006]{Sch06}). The value $Q \simeq 0.8$
separates radial clustering ($Q > 0.8$) from fractal type
clustering ($Q < 0.8$).
\begin{figure}[t]
\begin{center}
\resizebox{\hsize}{!}{\includegraphics{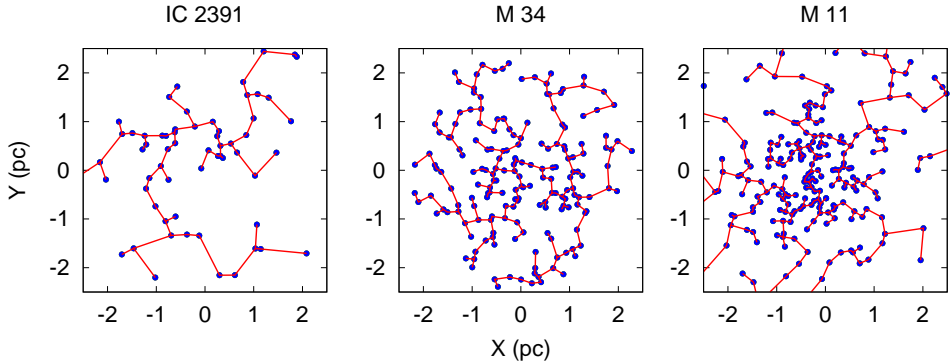}}
\caption{The minimum spanning tree is the set of straight
lines connecting the points such that the sum of their lengths
is a minimum. Here we show minimum spanning trees for three
open clusters, from which we can calculate the structure parameter
$Q$. Star positions are indicated with blue circles and red
lines represent the tree. The value of $Q$ quantifies
the spatial distribution of stars. For IC~2391 the stars
are distributed following an irregular fractal pattern
($Q = 0.77 < 0.8$), for M~34 the stars are distributed
roughly homogeneously ($Q = 0.8$), and for M~11 the stars
follow a radial density profile ($Q = 1.02 > 0.8$).}
\label{mst}
\end{center}
\end{figure}
\item Finally, we calculated the correlation dimension ($D_c$) and
its associated uncertainty by using an algorithm which gives reliable
results (\cite[S\'anchez et al. 2007a]{San07a}, \cite[S\'anchez \& Alfaro
2008]{San08}).
\end{enumerate}

\section{Main results}

Table~\ref{data} summarizes the relevant data (ages and
distances were taken from the Webda database).
\begin{table}
\begin{center}
\caption{Properties of the clusters in the sample.}
\label{data}
\begin{tabular*}{\textwidth}{
l@{\extracolsep{\fill}}
c@{\extracolsep{\fill}}
r@{\extracolsep{\fill}}
r@{\extracolsep{\fill}}
c@{\extracolsep{\fill}}
r@{\extracolsep{\fill}}
c@{\extracolsep{\fill}}
c}
\hline
Name & $\log\ T$ & $D$ & $N_s$ & $R_c$ & $R_t$ & $Q$ & $D_c$ \\
\hline
IC~2391  & 7.661 &  175 &   62 & 1.46 &  2.65 & 0.77 & $1.74 \pm 0.20$ \\
M~11     & 8.302 & 1877 &  289 & 1.98 &  4.49 & 1.02 & ... \\
M~34     & 8.249 &  499 &  181 & 0.11 &  1.73 & 0.80 & $2.04 \pm 0.05$ \\
M~67     & 9.409 &  908 &  354 & 2.21 &  5.92 & 0.98 & ... \\
NGC~188  & 9.632 & 2047 & 1459 & 2.90 & 10.57 & 0.91 & ... \\
NGC~581  & 7.336 & 2194 &  526 & 1.38 & 11.86 & 0.76 & $1.79 \pm 0.06$ \\
NGC~1513 & 8.110 & 1320 &  156 & 1.55 &  7.73 & 0.72 & $1.82 \pm 0.09$ \\
NGC~1647 & 8.158 &  540 &  683 & 1.23 &  8.86 & 0.70 & $1.94 \pm 0.02$ \\
NGC~1817 & 8.612 & 1972 &  277 & 3.39 & 11.97 & 0.79 & $1.94 \pm 0.04$ \\
NGC~1960 & 7.468 & 1318 &  311 & 2.96 &  8.77 & 0.87 & ... \\
NGC~2194 & 8.515 & 3781 &  228 & 3.17 & 10.31 & 0.85 & ... \\
NGC~2548 & 8.557 &  769 &  168 & 2.61 &  9.16 & 0.90 & ... \\
NGC~4103 & 7.393 & 1632 &  799 & 0.72 & 10.74 & 0.78 & $1.85 \pm 0.04$ \\
NGC~4755 & 7.216 & 1976 &  196 & 1.11 &  3.50 & 0.94 & ... \\
NGC~5281 & 7.146 & 1108 &   80 & 0.62 &  2.44 & 0.84 & ... \\
NGC~6530 & 6.867 & 1330 &  145 & 1.43 &  7.47 & 0.67 & $1.74 \pm 0.09$ \\
\hline
\end{tabular*}
\end{center}
\vspace{1mm}
{\it Note:} $T$ = cluster age (Myr); $D$ = distance (pc);
$N_s$ = number of members; $R_c$ = core radius (pc); $R_t$
= tidal radius (pc); $Q$ = structure parameter; $D_c$ =
correlation dimension.
\end{table}
On average, stars in young clusters tend to be distributed
following clustered, fractal-like patterns ($Q < 0.8$),
whereas older clusters tend to exhibit radial star density
profiles ($Q > 0.8$). However, the statistical analysis
indicates that there is no significant correlation between
$Q$ and $\log (T)$. If instead we consider the variable
$T/R_t$, which is proportional to the cluster age measured
in crossing time units (assuming nearly the same typical
velocity dispersion for the open clusters), then a
significant correlation is observed (Fig.~\ref{correlaQ}).
Additionally, we observe significant correlations
(confidence levels above 96 \%) between $D_c$ and $T$
(cluster age) and also between $D_c$ and $T/R_t$ (age in
crossing time units) for those clusters with internal
substructure (Fig.~\ref{correlaDc}).

\section{Discussion}

Our results support the idea that stars in new-born cluster
likely follow the fractal patterns of their parent molecular
clouds, and that eventually evolve toward more centrally
concentrated structures (see \cite[Schemja \& Klessen
2006]{Sch06}; \cite[Schmeja et al. 2008]{Sch08},
\cite[2009]{Sch09}; \cite[S\'anchez et al.  2007a]{San07a},
\cite[2009]{San09}). However, this seems to be only an
overall trend. The very young cluster $\sigma$ Orionis
(age $\sim 3$ Myr) exhibits a radial density gradient
with $Q \simeq 0.88$ (\cite[Caballero 2008]{Cab08}).
On the other hand, Table~\ref{data} shows open clusters
as old as $\sim 100$ Myr that have not totally destroyed
their clumpy structure (for example, both NGC~1513 and
NGC~1647 have $Q \sim 0.7$). \cite[Goodwin \& Whitworth
(2004)]{Goo04} simulated the dynamical evolution of
young clusters and showed that the survival of the initial
substructure depends strongly on the initial velocity dispersion.
Fractal clusters with a low velocity dispersion tend to erase
their substructure rather quickly. However, if the velocity
dispersion is high, such that the cluster remains supported
against its own gravity or even expands, then significant 
levels of substructure can survive for several crossing
times. Thus, our results give some observational support
to \cite[Goodwin \& Whitworth's (2004)]{Goo04} simulations.

\begin{figure}[t]
\begin{minipage}[t]{0.48\linewidth}
\includegraphics[scale=0.52]{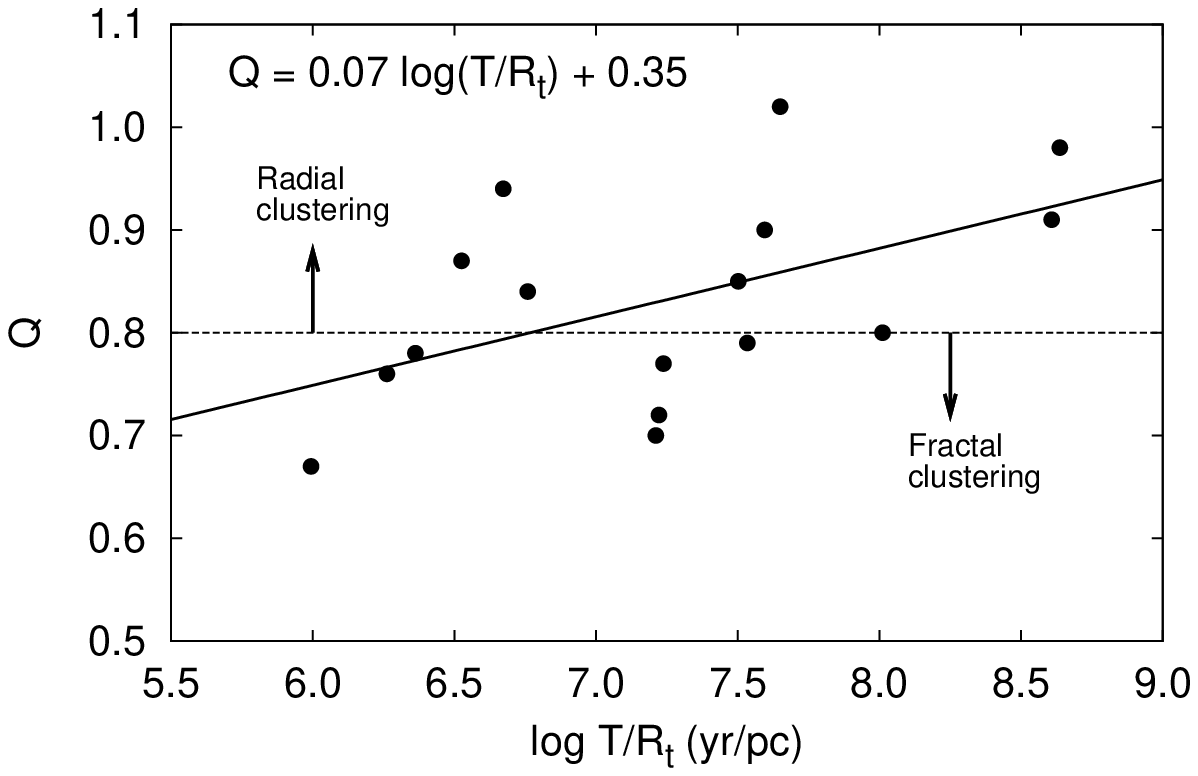}
\caption{Structure parameter $Q$ as a function of the
logarithm of age divided by the tidal radius, which is
nearly proportional to age in crossing times units. The
dashed line at $Q=0.8$ roughly separates radial from
fractal clustering. The best linear fit (equation at
the top) is represented by a solid line.}
\label{correlaQ}
\end{minipage}
\hfill
\begin{minipage}[t]{0.48\linewidth}
\includegraphics[scale=0.52]{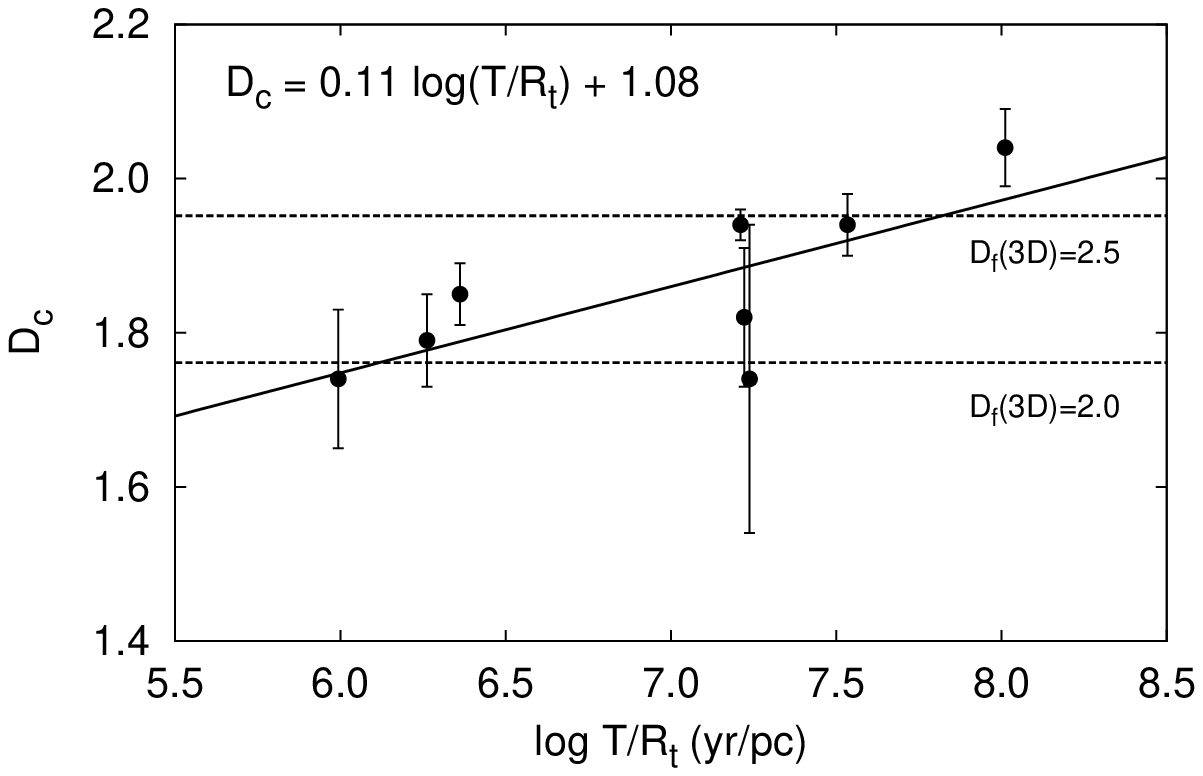}
\caption{Calculated correlation dimension as a function
of age (in crossing time units). The best linear fit
(equation at the top) is represented by a solid line.
As reference, horizontal dashed lines indicate the
values corresponding to three-dimensional distributions
with fractal dimensions of $D_f=2.0$ and $2.5$.}
\label{correlaDc}
\end{minipage}
\end{figure}

From Fig.~\ref{correlaDc}, we can see that clusters with
the smallest correlation dimensions ($D_c = 1.74$) would
have three-dimensional fractal dimensions around $D_f
\sim 2.0$ (estimated from previous papers, see e.g. Fig.~1
in \cite[S\'anchez \& Alfaro 2008]{San08}). This is a very
interesting result because this value is considerably smaller
than the average value estimated for galactic molecular clouds
in recent studies, which is $D_f \simeq 2.6-2.7$ (\cite[S\'anchez
et al. 2005]{San05}, \cite[S\'anchez et al. 2007b]{San07b}).
Young, new-born stars probably will reflect the conditions
of the interstellar medium from which they were formed.
Therefore, a group of stars born from the same cloud,
i.e. born at almost the same place and time, should have
a fractal dimension similar to that of the parent cloud.
If the fractal dimension of the interstellar medium has a
nearly universal value around 2.6-2.7, then how can some
clusters exhibit such small fractal dimension values?
Perhaps some clusters may develop some kind of substructure
starting from an initially more homogeneous state. This
possibility has been confirmed in numerical simulations
(\cite[Goodwin \& Whitworth 2004]{Goo04}), although some
coherence in the initial velocity dispersion is required.
Another explanation is that this difference is a consequence
of a more clustered distribution of the densest gas from which
stars form at the smallest spatial scales in the molecular
cloud complexes, according to a multifractal scenario
(\cite[Chappell \& Scalo 2001]{Cha01}). Maybe the star
formation process itself modifies in some (unknown) way 
the underlying geometry generating distributions of stars
that can be very different from the distribution of gas in
the parental clouds. Finally, one possibility is that the
fractal dimension of the interstellar medium in the Galaxy
does not have a universal value and therefore some regions
form stars distributed following more clustered patterns.
There is no a priori reason for assuming that $D_f$ has
nearly the same value everywhere in the Galaxy independently
of either the dominant physical processes or environmental
variables. Recent simulations of supersonic isothermal
turbulence done by \cite[Federrath et al. (2009)]{Fed09}
showed that compressive forcing yields fractal dimension
values for the interstellar medium significantly smaller
($D_f \sim 2.3$) compared to solenoidal forcing ($D_f \sim
2.6$). Thus, $D_f$ could be very different from region
to region in the Galaxy depending on the main physical
processes driving the turbulence. At least at galactic
scales, it has been shown that there are significant
differences in the fractal dimension of the distribution
of star forming sites among the galaxies, contrary to the
universal picture previously claimed in the literature
(\cite[see S\'anchez \& Alfaro 2008]{San08}). So that the
possibility of a non-universal fractal dimension for the
interstellar medium in the Galaxy cannot, in principle,
be ruled out.

\end{document}